\documentstyle[aps,prl,floats,twocolumn]{revtex}
\begin{document}
\draft
\twocolumn[\hsize\textwidth\columnwidth\hsize\csname @twocolumnfalse\endcsname
\title{Interactions and  scaling in a disordered two-dimensional metal}
\author{Sudip Chakravarty$^1$, Lan Yin$^1$, Elihu Abrahams$^{2}$}
\address{$^1$Department of Physics and Astronomy, University of California
Los Angeles\\
Los Angeles, CA 90095-1547\\
$^2$Serin Physics Laboratory, Rutgers University,
Piscataway, NJ 08855-0849\\}
\date{\today}
\maketitle
\begin{abstract}

We show that a non-Fermi liquid state of interacting electrons in two dimensions
 is stable in the presence of disorder and is a perfect conductor, provided the
interactions are sufficiently strong. Otherwise, the disorder leads to localization as in the case of non-interacting
electrons.  This conclusion is established by examining the replica field
theory in the weak disorder limit, but in the presence of arbitrary
electron-electron interaction. Thus, a disordered two-dimensional metal is a
perfect metal, but not a Fermi liquid.
\end{abstract}
]
A number of recent experiments \cite{exps} on the two-dimensional electron system in
semiconductor devices have revealed the existence of a low temperature
metal-insulator transition as the electron density is varied. In the 1980's,
similar experiments appeared to confirm the prediction that the two-dimensional
disordered electron system would have no metallic states\cite{g4}. What
distinguishes the new samples from those studied earlier is that the electron
density is extremely low, so that the Coulomb interaction energy is larger
than the Fermi energy.

The combined effects of disorder and electron-electron interaction were studied
by renormalization group (RG) methods \cite{amf,bk} and there were indications
that one effect of the interaction might be to stabilize a metallic state in two
dimensions. However, the nature of the metallic state was unknown, and, in any
case, the conclusion was not very definite since the RG flows went to a strong
coupling regime beyond the range of validity of the RG equations. Therefore, the
possibility of a two-dimensional metallic state remained an open issue from the
theoretical point of view.

A scaling analysis of the recent experiments has been made\cite{Dob}, and, based
on it, it was argued that any disordered two-dimensional
metal is a perfect metal but not likely to be a Fermi liquid.
The reason is that if the interaction is turned off, the electrons will
localize\cite{lr}, and the localized state has no resemblance to a Fermi gas.
This  motivated us to consider a two-dimensional non-Fermi liquid state and to
study the effects of impurities.

As a minimal specification of a non-Fermi liquid, the
retarded single particle Green's function must not contain a
quasiparticle pole when analytically continued to the lower half plane, but a
branch point. This leads to a spectral function  satisfying the
homogeneity relation \cite{Yin,footnote1}
\begin{equation}
A(\Lambda^{y_1}k,\Lambda^{y_2}\omega)= \Lambda^{y_A}A(k,\omega),
\end{equation}
in the asymptotically low energy limit, where $y_1$, $y_2$, and $y_A$ are the 
exponents defining the
universality class of
the non-Fermi liquid. Only the set of exponents
$y_1=1$, $y_2=1$, and $y_A=-1$ represents a Fermi liquid
for which the branch points collapse into simple poles. Here the momentum is
measured with respect to $k_F$ and the
frequency is measured with respect to the Fermi energy.

The above spectral
function will be assumed to contain a kinematic form factor
of zero scale dimension, which is $\theta(\omega^2-v_F^2 k^2)$. The
rationale is as follows: if dissipation is due to the decay of an electron
coupled to particle-hole pairs, then, for $\omega > 0$, $\omega$ has to be
greater than $v_F|k|$. Similarly, because in the ground state of a non-Fermi
liquid, particles are present both
above and below  the Fermi sea,  $\omega$ must be less than $-v_F|k|$ for
negative frequencies. The presence of the
$\theta$-function leads to a density of states that vanishes
at the Fermi energy. In Ref.~\cite{Yin}, another choice was made for which
the density of states
remains finite at the Fermi surface.

A spectral function with non-trivial exponents is necessary, but it does not
fully specify  a non-Fermi liquid. For example, it does not contain
spin-charge separation, which requires separate singularities for spin and
charge excitations.
Moreover, new exponents may have to be introduced for the scaling of composite
operators. It is,
however, our intention to see what can be learned from this minimal
specification of a non-Fermi liquid.

For explicit calculations we use the simple model:
\begin{equation}
A(\omega,k)\propto {1\over
\omega_c^{\alpha}|\omega-v_F k|^{1-\alpha}}\theta(\omega^2-v_F^2 k^2),
\label{spec}
\end{equation}
where $\omega_c$ is a microscopic high frequency scale proportional to the
inverse of the noninteracting density of states $\nu$.  The exponent $\alpha$ has to be positive to
satisfy the analyticity properties required of the single particle Green's function. It is also worth
noting that the singular part of the spectral function alone will not satisfy any sum rules.

The grassmannian field theory for the localization problem was first set up by
Efetov, Larkin and Khmel'nitskii\cite{elk} and extended by Finkel'stein\cite{amf} to the interacting
case. We follow Ref. \cite{amf}. The random impurity potential is spatially uncorrelated and has a
white-noise distribution with zero mean so that
$
\langle V(r) V(r')\rangle =(1/ 2\pi \nu\tau)\delta^{(d)}(r-r'),
$
The disorder average is carried out on the replicated partition function
$Z_N$, where $N$ is the number of replicas. It leads to  the action
\begin{equation}
S=\sum_{\alpha} [S^{\alpha}_0+S^{\alpha}_{\rm e-e}]+R_0\int d^dr\left({\overline
\psi}\psi\right)^2,
\end{equation}
where
$
\left({\overline \psi}\psi\right)=\sum_{n,\alpha}{\overline
\psi}_n^{\alpha}\psi_n^{\alpha}.
$
and $R_0=1/4\pi \nu\tau$. The Grassmann variables
$\{{\overline \psi}^{\alpha}_n(r),\psi^{\alpha}_n(r)\}$ carry Matsubara
index $n$ ($\omega_n=\pi(2n+1)T$); the spin and the replica
indices are lumped together in $\alpha$. Here
$S^{\alpha}_0$ is the action of free fermions. It and the electron-electron
interaction term $S^{\alpha}_{\rm e-e}$ is diagonal in replica indices.
Replica mixing
takes place only in the impurity-induced interaction term.

If the electron-electron interaction leads to a Fermi liquid, the
impurity induced interaction
is relevant by power counting. Let us
rewrite it  as
\begin{equation}
R_0
\sum_{\alpha,\beta}\int d^dr d\tau
d\tau'\overline{\psi}^{\alpha}(r,\tau)\psi^{\alpha}(r,\tau)
\overline{\psi}^{\beta}(r,\tau')\psi^{\beta}(r,\tau')\label{quartic}
\end{equation}
Under the scale transformations $k\to \Lambda k$ and $\omega \to \Lambda
\omega$, the dimension  of
the fermion variables obtained from the free fermion action is \cite{Shankar}
$
[\psi^{\alpha}(k,\omega)]\sim \Lambda^{-({3/2})}
$
so that $
[\psi^{\alpha}(r,\tau)]\sim \Lambda^{(1/2)}.
$
Note that spatial dimensionality, $d$, does not enter into the scale dimensions, because the scaling is
in a direction normal to the Fermi surface; the transverse $(d-1)$ directions act as internal degrees
of freedom.  Then, from Eq.~(\ref{quartic}), the impurity-induced interaction scales as
$
R(l)=R_0e^{l},
$
where $\Lambda^{-1}
= e^l$ and is thus relevant. Loop corrections show that either the system flows
to a diffusive metal fixed point (a
possibility that exists for $d> 2$) or to the  Anderson localized phase. The
possibility of a
pure (ballistic) Fermi liquid phase does not exist because disorder is always a
relevant
operator \cite{Nayak}.

In contrast, let us suppose that $S_0^{\alpha}$ + $S_{\rm e-e}^{\alpha}$ leads
to a non-Fermi
liquid state. Consider the simple model in Eq.~\ref{spec}. Then
it follows that
$
\left[\psi^{\alpha}(k,\omega)\right] \sim \Lambda^{-({3-\alpha\over 2})}
$,
and
$
\left[\psi^{\alpha}(r,\tau)\right] \sim \Lambda^{1+\alpha\over 2}.
$
Consequently,
$
R(l)=R_0e^{(1-2\alpha)l}
$.
Now the impurity-induced interaction is {\em irrelevant} for $\alpha > 1/2$ and
 relevant for
$\alpha < 1/2$. For $\alpha>1/2$,
the disordered system is described by a non-Fermi liquid fixed
point\cite{Nayak}.  For
$\alpha < 1/2$, the system has a choice between a diffusive metal or a Anderson
localized phase.

To explore this situation further, we rewrite the action in terms of slow
variables by isolating the regions of small total momentum in the replica-mixed
particle-hole channel
\cite{amf}. We define a matrix $B$ in the replica, spin and Matsubara indices
by
$R_0\left({\overline \psi}\psi\right)^2=-{1\over 2}{\rm Tr} B^2(r)$.  The
Fourier transform of $B$ is
\begin{equation}
B^{\alpha\beta}_{nn'}(q)={1\over \sqrt{2\pi\nu \tau \Omega}}\sum_k
\overline{\psi}^{\alpha}_n(k)\psi^{\beta}_{n'}(k+q),
\end{equation}
where $\Omega$ is the sample volume. Introducing a hermitian matrix field $Q(r)$
to decouple this term
\cite{amf}, we find,
using a trick due to Schwinger \cite{schw}, which is valid even when the action
is not quadratic in Grassmann
variables, that
\begin{equation}
\overline{Z}_N \sim \int {\cal D}\{Q\} e^S =  \int {\cal D}\{Q\} e^{-{1\over 2}\int d^d r\big[
{\rm Tr} Q^2 + 2i \int_0^1
d\lambda \langle {\rm Tr} QB\rangle_{\lambda}\big]},
\end{equation}
where we have taken the replica limit $N\to 0$ in a prefactor, and
\begin{eqnarray}
\langle &&{\rm Tr} QB\rangle_{\lambda}\nonumber\\
&&=i\frac{d}{d\lambda} \log \left[ \int {\cal
D}\{\overline{\psi},\psi\} e^{ \sum_{\alpha}
(S^{\alpha}_0+S^{\alpha}_{\rm e-e})-i\lambda\int d^d r {\rm Tr} QB }\right].
\end{eqnarray}

From Feynman diagrams, one can see
that electron-electron interaction cannot dress the $QB$ vertices when the
replica-off-diagonal components of
$Q$ are involved. However, if $Q$ is diagonal, the vertices can receive
corrections from electron-electron interaction. We shall neglect the vertex
corrections for the replica-diagonal background field $Q^{\alpha\alpha}$.
This approximation breaks gauge
invariance, but has little
effect on the localization problem, which crucially depends on the off-diagonal
$Q$'s. With this
approximation \cite{footnote2}, the quantity $\langle {\rm Tr}
QB\rangle_{\lambda}$ can be calculated from the matrix
Green's function $G^{\alpha\beta}_{mn}(r,r';\lambda)$, which satisfies the Dyson equation
\begin{eqnarray}
G(r,r';\lambda)&=&G(r-r';0)\nonumber \\
&-&{i\lambda \sqrt{2R_0}}\int d^d r_1
G(r-r_1;0)Q(r_1)G(r_1,r';\lambda).\nonumber \\
& &
\end{eqnarray}
The Green's function $G(r-r',i\omega_m;0)\delta_{mn}\delta_{\alpha\beta}$ is the fully interacting
  one, but in the absence of the matrix field
$Q(r)$. It is translationally invariant and diagonal in Matsubara and
replica indices. As a result of our approximation, $G(r-r';0)$ fully determines
$G(r,r';\lambda)$ as a
functional of $Q(r)$.

Consider the saddle point solution for which $Q$
is a constant diagonal matrix, that is,
$Q(r)=\overline{Q}(i\omega_n)\delta_{mn}\delta_{\alpha\beta}$ \cite{footnote1}. Then, the self-consistent
equation for $\overline{Q}$ is
\begin{equation}
\overline{Q}(i\omega_n)={i\over \Omega}\sum_k{\sqrt{2R_0}G(k,i\omega_n;0)\over 1+i \overline{Q}(i\omega_n)\sqrt{2R_0}
G(k,i\omega_n;0)}.
\end{equation}
In the limit $(\nu/\tau)\ll 1$, we can write, to leading order,
\begin{equation}
\overline{Q}(i\omega_n) \approx {i\sqrt{2R_0}\over \Omega}\sum_k G(k,i\omega_n;0)
\end{equation}
Using the explicit form of the spectral function given earlier, it is possible
 to show that
\begin{equation}
\overline{Q}(i\omega_n)=A(\alpha)\sqrt{\pi\nu\over 2\tau}{1\over
\cos(\alpha\pi/2)}|\omega_n/\omega_c| ^{\alpha}
{\rm sgn}(\omega_n),
\end{equation}
where $A(\alpha)$ is a function of order unity.  As the next iteration, one
may substitute this
expression for $\overline{Q}$ in the exact saddle point equation to estimate the
correction. For $\alpha
> 1/2$, it is readily
seen that the correction vanishes in the limit of zero frequency. For
$\alpha < 1/2$, the correction tends to infinity so that Eq.\ (11) cannot be a correct solution.

For $\alpha < 1/2$ and not too close to it, the saddle point value of $\overline{Q}$ can
be calculated for the model spectral
function from Eq.\ (9) without the approximation of Eq.\ (10). 
For the frequency tending to zero, it is
\begin{equation}
\overline{Q}(i\omega_n)=B(\alpha)\left({\pi\nu\over 2\tau}\right)^{1\over 2(1-2\alpha)}{\rm sgn}
(\omega_n) \equiv q {\rm sgn}(\omega_n),
\label{saddle}
\end{equation}
where $B(\alpha)$ is a function  of order unity, and we have neglected the weak frequency dependence.
The saddle point solution  for $\alpha < 1/2$ requires careful treatment of the cutoff. In
the limit $\alpha \to 1/2$, the presence of the $\theta$-function cutoff in the
non-Fermi liquid spectral function cannot be ignored, while for $\alpha \to 0$
it can be. If this crossover is treated correctly, one finds that as $\alpha\to
1/2$,
\begin{equation}
\overline{Q}(i\omega_n)=\sqrt{2\pi \tau|\omega_n|}\left[(\exp({\pi\tau\over \nu})-1)^2-({2\tau\over
\nu})^2\right]^{-{1\over 4}}{\rm sgn}(\omega_n).
\end{equation}
This solution smoothly joins the solution given in Eq.\ (12) at $\alpha=1/2$.

To summarize: In the non-disordered case, the density of
states for the non-Fermi liquid vanishes at zero frequency for all $\alpha$. In
the presence of disorder, this quantity is considered as an order parameter, and
the non-Fermi liquid behavior survives if $\alpha > 1/2$, that is, when the the order parameter vanishes. If, on
the other hand, $\alpha < 1/2$, the symmetry is broken, and there is a non-zero density of
states in the disordered situation.

Next, we investigate the fluctuation modes about the mean-field solution. We consider a  small deviation $\delta
Q=(Q(r)-\overline{Q})$.
It is useful to shift the variable further\cite{amf} by shifting $\overline{Q}$ to
${\overline Q}+H$, where
\begin{equation}
H(i\omega_n)=i\sqrt{2\pi\nu\tau}\left[G^{-1}(k,i\omega_n;0)-G^{-1}(k,0;0)
\right]I,
\end{equation}
where $I$ is the unit matrix in replica indices. Then it is not difficult to see that
\begin{equation}
S=-{\cal F}({\overline Q}+H)-{1\over 2}\int d^dr\left[{\rm
Tr}(\delta\tilde{Q})^2
+ 2{\rm
Tr}H{\delta\tilde Q}\right]+\delta S,
\end{equation}
where ${\delta\tilde Q} = Q - {\overline Q} - H$. The quadratic part of $\delta S$ is given by
\begin{equation}
\delta S_2=R_0\int {d^d k\over
(2\pi)^d}\delta {\tilde Q}_{mn}^{\alpha\beta}(k)\delta{\tilde
Q}_{nm}^{\beta\alpha}(-k){\tilde
F}_{mn}(k),
\end{equation}
with
\begin{equation}
{\tilde F}_{mn}(k)=\int {d^d p\over (2\pi)^d}{\tilde G}(p,i\omega_m){\tilde G}(p-k,i\omega_n).
\end{equation}
Here
\begin{equation}
{\tilde G}(k,i\omega_n)= {1\over G^{-1}(k,0;0)I+i 
{\overline Q}(i\omega_n)\sqrt{2R_0}}.
\end{equation}
Note  that $H$ plays the role
of a symmetry breaking
field, which in the noninteracting case collapses to
$\sqrt{2\pi\nu\tau} \omega_n I$. The shift,
however, has allowed us to recast the expansion in terms of $\tilde G$
which is invariant under
arbitrary homogeneous unitary rotations of the saddle point.

The $\delta{\tilde Q}$ fluctuation modes will be classified as longitudinal when $mn>0$ and transverse when
$mn<0$. In the Fermi liquid case, the latter are the usual diffusion modes.
The
longitudinal modes are massive as in the non-interacting model \cite{elk}. The zero momentum limit of the 
transverse part of
${\tilde F}_{mn}(0)$ for $\alpha < 1/2$ is given by
\begin{equation}
{\tilde F}_{mn}^T(0)={1 \over \Omega} \sum_k {G^2(k,0;0) \over {1+ 2 R_0 q^2 G^2
(k,0;0)}},
\end{equation}
where $q$ is defined in 
Eq.~(\ref{saddle}). One can verify using the saddle point equation that
${\tilde F}_{mn}^T(0)=(1/ 2R_0)$.
This leads to a complete cancellation between the quadratic term $-{1\over 2}\int
d^dr {\rm
Tr}(\delta\tilde{Q})^2$ and $\delta S_2$ in the fluctuation action, Eq.\ (16). The resulting theory 
for the transverse modes is
massless, and the remaining
quadratic term in the action 
in the long wavelength limit is
\begin{equation}
\delta S_2= -{\pi \nu D(\alpha)\over 4}\int {d^d k\over
(2\pi)^d}k^2\delta {\tilde Q}_{mn}^{\alpha\beta}(k)\delta{\tilde
Q}_{nm}^{\beta\alpha}(-k),
\end{equation}
where the bare diffusion constant at the scale of the mean free path is
renormalized by the
electron-electron interaction and is ($\alpha < 1/2$ and not too close)
\begin{equation}
D(\alpha)=(1-\alpha)\left[{\pi\over B(\alpha)}\right]^{2\alpha\over 1-\alpha}
\left[2\tau\over
\pi\nu\right]^{2\alpha\over 1-2\alpha}D,
\end{equation}
where $D=v_F^2 \tau/d$ is the usual diffusion constant. In this equation we
have also
normalized the saddle point solution such that $\overline{Q}^2=I$.  As in the
non-interacting
problem, the resulting theory can now be  cast  in the language of a matrix
nonlinear
$\sigma$-model. It is well-known that all states are localized for $d\le 2$. The
 only difference is that the bare diffusion constant $D(\alpha)$ is enhanced
over the
non-interacting value $D$. 

As in the corresponding non-interacting problem, the mass term cancels in the transverse channel and the
diffusion propagator diverges as $k^{-2}$ at zero frequency, for  $\alpha < 1/2$. These are properties of the
Goldstone modes in the transverse replica channel and follow as long as the
continuous symmetry in the $Q$-field theory is broken, which is indeed the case for $\alpha < 1/2$; they do not depend critically on the
assumptions made in the longitudinal channel. Thus, the neglect of the vertex correction in the longitudinal channel does not spoil this picture.

For $\alpha>1/2$, ${\overline Q}(i\omega_n)$ tends to zero
as the frequency tends to zero, and the transverse part of the mass term
does not cancel against the
transverse part of
$\delta S_2$.
Our analysis shows that impurity interactions are irrelevant for
$\alpha
> 1/2$, and  the theory  corresponds to the  non-Fermi liquid fixed point
Hamiltonian, minimally specified by the scaling property of the one-particle
spectral function. It is preferable then to return to the original fermion language.
Such a system should be a perfect metal.

It is important to make a number of qualifying remarks.

(1) For $\alpha > 1/2$, any attempt to
expand in powers of the exponent $\alpha$ will  lead to numerous
logarithmic divergences, and the
theory cannot be cast in the language of  a Fermi liquid
theory.  For $\alpha < 1/2$, the electrons are localized
in $d\le 2$, with the interacting bare diffusion constant enhanced over the
non-interacting value.
Thus, once again, the language of Fermi liquid theory is not meaningful.

(2)The model spectral function considered here does not contain spin charge
separation. From a
simple model reflecting spin charge
separation\cite{Yin}, it is possible to see that our
qualitative conclusions regarding broken symmetry remain unchanged. It is not apparent,
however, that that is the correct model for spin-charge separation.
In fact, the entropy producing process in which an accelerated
electron breaks
up into a spin and a charge excitation is missing\cite{Anderson}. 
It is also possible that  spin-charge separation will affect the behavior in a magnetic field.
For example, in a
two-dimensional system, a parallel magnetic field can couple to the spinon
excitations, producing a
gap in the spectrum, thereby destroying the non-Fermi liquid state. This, in turn, will
lead to
a localized insulating state as seen in experiments\cite{Sarachik}.

(3) The results obtained here are  only superficially similar to those of
the one-dimensional problem\cite{giam}.  In
one-dimensional theories, the anomalous dimension of the density-density
correlation function cannot
be simply obtained from the anomalous dimension of the one-particle Green's
function. As remarked earlier, the operator product expansion of composite operators
contain further
singularities.  Such possibilities were purposely avoided to reduce the
complexity of the
theory.  Nonetheless, we have demonstrated that
the breaking of the
single most important assumption of the Fermi liquid theory, namely the
existence of the
quasiparticles,
leads to 
dramatically different behavior with respect to the localization properties of
the interacting
electron system. Moreover, it appears that there are experimentally relevant regimes in two dimensions, which
show no hint of instabilities that could arise from $2k_F$ vertex corrections. Thus, the neglect of vertex corrections
in two dimensions may have a well-defined range of validity.

It is a pleasure to acknowledge important discussions with A. Balatsky, D. Belitz, T. Kirkpatrick, A. Ludwig,
 S. Sachdev, R. Shankar, H. Schulz, G. Zimanyi, S. C. Zhang, and especially C.
Nayak. A part of this work was carried out
at the Aspen Center for Physics. This work was supported by grants from the
National Science
Foundation: DMR-9531575 (S.C. and L.Y.) and DMR 9632294 (E.A.).


\begin{references}
\bibitem{footnote1}It is sufficient to consider a translationally invariant
definition as
the replicated field theory is indeed translationally invariant.
\bibitem{footnote2} This approximation fails in one
dimension where such vertex corrections are governed by composite operators which are actually
singular at momentum transfer $2k_F$.
\end{references}
\end{document}